\documentclass{WileyMSP-template}
\usepackage{url}
\usepackage{physics}

\begin{document}
\pagestyle{fancy}
\rhead{\includegraphics[width=2.5cm]{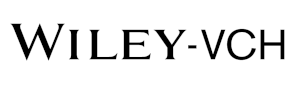}}

\title{Acoustic Manipulation of Tangible Janus Icons on Liquid Droplets}

\maketitle

\author{Yusuke Koroyasu}
\author{Yoichi Ochiai}
\author{Tatsuki Fushimi*}

\begin{affiliations}
Yusuke Koroyasu\\
School of Informatics, College of Media Arts, Science and Technology, University of Tsukuba, Kasuga Campus Kasuga 1-2, Tsukuba, Ibaraki, 305-8550, Japan\\
Email Address: koroyu@digitalnature.slis.tsukuba.ac.jp\\

Prof. Yoichi Ochiai\\
R\&D Center for Digital Nature, University of Tsukuba, Tsukuba, 305-8550, Ibaraki, Japan\\
Institute of Library, Information and Media Science, University of Tsukuba, Tsukuba, 305-8550, Ibaraki, Japan\\
Tsukuba Institute for Advanced Research (TIAR), University of Tsukuba, 1-1-1 Tennodai, Tsukuba, 305-8577, Ibaraki, Japan\\
Pixie Dust Technologies, Inc., Chuo-ku, 104-0028, Tokyo, Japan
Email Address: wizard@slis.tsukuba.ac.jp\\

Prof. Tatsuki Fushimi\\
Institute of Library, Information and Media Science, University of Tsukuba, Tsukuba, 305-8550, Ibaraki, Japan\\
R\&D Center for Digital Nature, University of Tsukuba, Tsukuba, 305-8550, Ibaraki, Japan\\
Tsukuba Institute for Advanced Research (TIAR), University of Tsukuba, 1-1-1 Tennodai, Tsukuba, 305-8577, Ibaraki, Japan\\
Email Address: tatsuki@levitation.engineer
\end{affiliations}



\begin{abstract}
Interfaces that couple digital information with physical matter enable computation to be expressed through tangible motion and touch, yet typically rely on embedded actuators, rigid mechanisms, or enclosed environments. Consequently, contactless manipulation and interaction with centimeter-scale tangible elements in open settings remain difficult to achieve. Here, we present PolygonWave, a solid--fluid acoustic interface that enables transport and tangible interaction by coupling airborne ultrasound with liquid-mediated support. The system employs lightweight Janus icons with asymmetric wettability: a superhydrophobic upper surface permits dry touch interaction, while a hydrophilic lower surface couples to a water droplet resting on a superhydrophobic mesh. Focused acoustic fields generated by a 256-element phased array induce lateral forces, enabling programmable motion without mechanical contact. Systematic characterization demonstrates transport of payloads up to 525~mg across variations in icon size, droplet volume, and applied load. Beyond translation, the liquid layer functions as a reconfigurable mechanical element, enabling button-like input with self-recovery and resonance-driven vibro-visual feedback, exhibiting a peak response near 22~Hz for 200~\textmu L droplets. Liquid-mediated acoustic coupling provides a unified mechanism for mechanically expressive, touch-accessible tangible interfaces bridging acoustics, soft matter physics, and physical human--computer interaction.
\end{abstract}


\section{Introduction}
Recent work on physical AI has emphasized the integration of perception, computation, and actuation to enable systems that interact directly with the physical world. While sensing and computation have advanced rapidly, actuation remains a limiting factor. Enabling actuation without physical contact would further extend the embedding of AI into physical environments, allowing computational intent to be expressed through physical manipulation without direct mechanical coupling.

A wide range of research communities have investigated approaches to contactless manipulation. In materials science, engineered wettability, capillarity, and architected surfaces have been used to transport liquids over surfaces \cite{Chaudhury1992,Chen2016,Li2017,Dudukovic2021,McCracken2020}. In microfluidics, droplet control is achieved through programmable platforms such as electrowetting-on-dielectric (EWOD) \cite{lee1998liquid, Nelson2012} and field-based actuation approaches including optical, acoustic, and electrostatic techniques \cite{Yan2021, Mi2021, Li2020, Koroyasu2023}, enabling precise transport, positioning, and manipulation of liquid volumes. Beyond droplet routing, liquids can also serve as load-bearing and motion-transmitting mechanical elements. For example, droplets have been used as active components in micro-conveyor systems \cite{Moon2006}, as lubricated supports in droplet-supported micromotors \cite{He2019}, and as tunable liquid bearings with controllable stiffness \cite{Ni2018}.

Acoustic manipulation~\cite{Foresti2013a,Andrade2018,Inoue2019,Marshall2012} and pneumatic approaches~\cite{Laurent2015,Matignon2010} provide non-contact actuation in air and have enabled transport of liquids and solids without direct mechanical contact. More recently, airborne ultrasound has been used to achieve two-dimensional transport of centimeter-scale objects through specific geometric features~\cite{Arakawa2024}, and to levitate larger and heavier objects using time-averaged acoustic trapping fields~\cite{Hirai2025}. In parallel, human–computer interaction~\cite{Lee2011,umapathi2018programmable,Lin2023,Wang2024} and robotics~\cite{Rubenstein2014,An2023,Yang2025} have introduced actuated tangibles, levitated interfaces, and cooperative object transport using embedded actuation (for e.g.~motors) or external agents.

Despite this diversity of approaches, contactless manipulation of centimeter-scale tangible elements with arbitrary shape in open environments remains technically challenging. A central difficulty is that the physical mechanisms underlying existing methods do not scale uniformly beyond small droplets or objects, which limits how effectively actuation can be extended to macroscopic artifacts in everyday settings. For example, in acoustic manipulation, radiation forces scale approximately with object cross-sectional area, whereas gravitational loading scales with volume, leading to a progressive reduction in the available force margin for levitation and controlled motion as object size increases~\cite{Fushimi2020}. While recent acoustic systems achieve precise control of object motion~\cite{Arakawa2024,Hirai2025}, they remain poorly suited for sustained tangible interaction. Similarly, air-cushion and pneumatic levitation systems can support larger planar payloads~\cite{Laurent2015,Matignon2010}, but their reliance on air films and distributed jets primarily enables in-plane motion and limits their applicability to touch-based interaction. 

\begin{figure}
\centering
  \includegraphics[width=\linewidth]{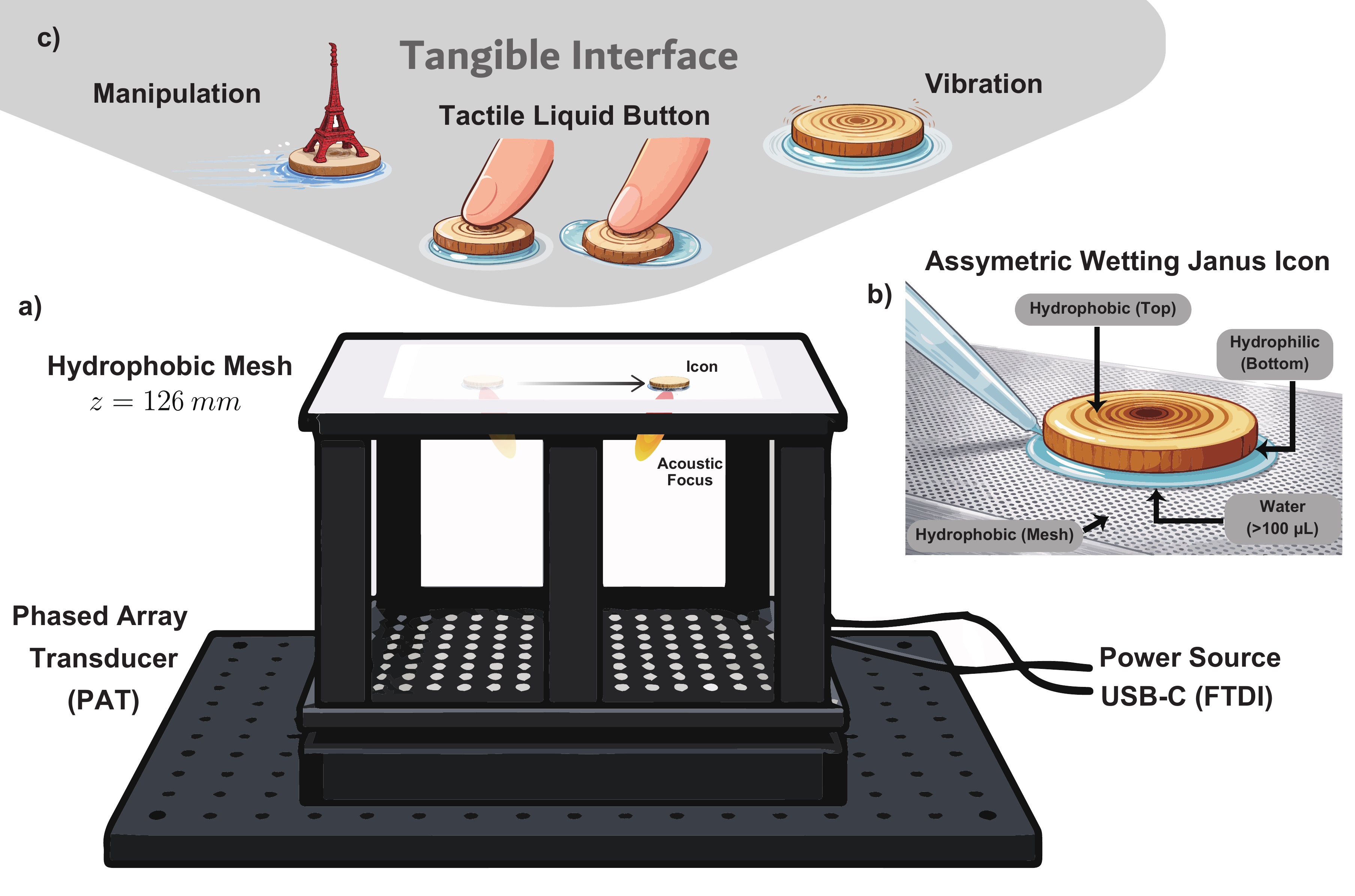}
  \caption{Overview of the solid--fluid acoustic tangible interface. 
(a) System configuration consisting of a phased-array transducer (PAT) positioned beneath a superhydrophobic mesh, which supports a liquid layer and enables contactless translation of Janus icons through programmable acoustic focusing. 
(b) Structure of the asymmetric wetting Janus icon, featuring a hydrophobic top surface for dry touch interaction and a hydrophilic bottom surface that couples to the supporting liquid layer. 
(c) Interaction modalities enabled by the system, including horizontal manipulation of mounted objects, tactile button input, and vibro-visual feedback.}
  \label{fig:conceptual_figure}
\end{figure}

Building on our prior mesh-based acoustic microfluidic interface~\cite{Koroyasu2023}, we present PolygonWave, a hybrid solid–fluid interface that couples liquid-mediated support with air-transmitted ultrasound for contactless manipulation and interaction with centimeter-scale tangible elements (Fig.\ref{fig:conceptual_figure}a). The platform is built around Janus icons—lightweight, touchable carriers whose opposing faces exhibit contrasting wettability (hydrophilic and superhydrophobic) (Fig.\ref{fig:conceptual_figure}b). By supporting tangible icons on liquid interfaces rather than fully levitating them, the system exploits capillary coupling to transmit forces with reduced actuation requirements, while maintaining a dry interaction surface. This enables open, touch-accessible manipulation without exposed electrodes or sustained optical excitation~\cite{Moon2006,Li2020,Yan2021}. 

As a result, transport can be achieved with reduced dependence on object geometry and increased tolerance to additional payloads, enabling stable manipulation of masses up to 525 mg using a single phased array of 256 ultrasonic transducers. Although this maximum payload is lower than those reported in large-scale acoustic systems\cite{Hirai2025,Arakawa2024}, it is achieved with a substantially more compact and simpler hardware configuration. Beyond transport, the liquid-supported icons function as mechanically active interface elements. The same liquid–acoustic coupling that supports loaded motion enables centimeter-scale tangibles to be pressed as compliant tactile buttons and driven into resonance for vibro-visual feedback (Fig.~\ref{fig:conceptual_figure}c), establishing a unified mechanism for manipulation and interaction. Together, these capabilities demonstrate how contactless actuation can be combined with stable, physically supported tangible interaction to strengthen the integration of computation and physical action in open environments.
\section{Results}

\subsection{Transport Capability Across Icon Size, Droplet Volume, and Load}
We first characterized the transport capability of the PolygonWave across combinations of icon diameter, droplet volume, and applied load. Transport arises from coupled interactions at the mesh–droplet and droplet–icon interfaces, and both the effective load supported by the droplet and the applied acoustic radiation force depend strongly on system geometry. As a result, performance is governed by multiple interrelated properties that are difficult to predict numerically, requiring accurate modeling of surface energy, experimental variability, and acoustic radiation forces. Rather than attempting continuous optimization over this high-dimensional and nonlinear parameter space, we evaluated transport behavior using a functional criterion: whether an icon could follow a prescribed circular trajectory under load.

All experiments were conducted using a single acoustic focus positioned 2 mm above the mesh surface, with the mesh located 126 mm from the transducer array. Details of the phased-array transducer system are provided in the Experimental Section (Phased Array Transducer). The operating principle by which a focused acoustic field generates an attractive traverse force is described in detail in our previous work~\cite{Koroyasu2023}. The fabrication and physical properties of the Janus icons and the superhydrophobic mesh are described in the Experimental Section (Physical Properties and Preparation of Icons; Physical Properties and Preparation of Mesh). For each trial, an icon was placed at a fixed starting position, and the droplet volume, icon size, and applied load were initialized. Droplet volume was adjusted incrementally using an electronic pipette (A\&D MPA-200), and deionized water (Koga Chemical Mfg.\ Co., Ltd.) was used for all experiments. When changing the applied load, the droplet was removed, the load was repositioned as close as possible to the geometric center of the icon, and water was reapplied. Three load conditions were tested: 135 mg (2.6 mm chrome-plated washer), 360 mg (2.6 mm chrome-plated nut), and 525 mg (7 mm steel disc). For each combination of droplet volume and load, the experiment was repeated three times.

Figure~2a summarizes the observed transport outcomes across this parameter space. Each bar represents the proportion of trials that fell into one of five outcome categories. Cases labeled as [V] Stable Trajectory Control (green) correspond to successful transport, in which the icon followed the prescribed circular trajectory without significant deviation. [IV] Partial Trajectory (blue) represents cases in which the icon was actuated and transported by ultrasound but deviated from the target trajectory during motion. These two categories indicate operational regimes in which transport is achievable in principle, and representative trajectories are provided in the Supplementary Material.

Three additional categories correspond to failure modes. [III] Lifted (No Translation) (pink) occurs when capillary forces are sufficient to support the weight of the icon and applied load, but the lateral acoustic radiation force is insufficient to induce horizontal motion. [II] Grounded (No Lift) (orange) appears when the applied load is too large or when the droplet volume is insufficient to fully occupy the underside of the icon, causing the icon to remain in contact with the mesh and experience increased friction, thereby preventing motion. Finally, [I] Flips (red) corresponds to cases in which the vertical acoustic force exceeds the combined weight, leading to overturning of the icon. Across all tested conditions, stable or partial transport was observed within a limited region of the parameter space, reflecting the competing requirements of sufficient lateral force for translation, adequate vertical support, and mechanical stability of the droplet--icon system.
\subsection{Selection of Operating Conditions for Subsequent Experiments}
Based on the parameter-space characterization in Figure~2a, icons with a diameter of 12~mm exhibited the highest proportion of stable trajectory control outcomes across multiple droplet volumes and load conditions. From a transport-oriented perspective, this configuration appears favorable. However, closer inspection reveals practical limitations that reduce its suitability for interactive use.

As shown in Figures2b–d, increasing the droplet volume on 12 mm icons leads to progressive vertical elongation of the droplet, which induces increased tilting of the icon during operation. Such tilting complicates the placement and retention of external loads on the icon and reduces its suitability for carrying mounted structures. For practical manipulation and interaction, it is therefore desirable for the icon surface to remain approximately horizontal throughout motion.

In addition, the small diameter of the 12~mm icons introduces limitations for direct human interaction. The average width of an adult human fingertip is approximately 13.3~mm \cite{kawachi2012hand}, based on the AIST hand dimension dataset of Japanese males and females ($N=530$; fingertip breadth of the index finger). As a result, manipulating smaller icons increases the likelihood of unintended contact with the exposed liquid. This elevates the risk of disturbing the droplet--icon interface and reduces usability in tangible interaction scenarios. Consequently, despite achieving a high proportion of stable transport outcomes, the 12~mm configuration was deemed unsuitable for further experiments.

In contrast, Figures~2e--g show that 15~mm icons remain approximately horizontal as the droplet volume is increased from 100~\textmu L to 300~\textmu L. The increased lateral extent of the icon distributes the droplet more evenly, suppressing vertical elongation and limiting tilt. This improved mechanical stability is advantageous for both sustained transport and direct user interaction.

Among the tested droplet volumes for the 15~mm icons, 100~\textmu L enabled stable transport but was close to the threshold for overturning, and occasional flipping was observed when the droplet was slightly mispositioned. A volume of 200~\textmu L provided greater vertical stability, although it predominantly resulted in partial trajectory outcomes under the simple single-focus actuation scheme used in this study. Nevertheless, this regime consistently supported loaded transport and offered improved robustness against tilting and handling disturbances.

Figure~2a further indicates that multiple regions of the parameter space support transport, even outside the stable trajectory control regime. For example, icons with a diameter of 18~mm were capable of transporting loads of up to 525~mg under certain droplet volumes. However, for these larger icons, insufficient liquid volume often led to grounding due to incomplete vertical support, whereas excessive acoustic forcing at low total mass increased the likelihood of flipping (i.e.~the larger surface area also enhances vertical acoustic radiation forces).

Taken together, these results indicate that practical operating conditions must balance transport performance, mechanical stability, and interaction usability. Based on this consideration, subsequent experiments were conducted using 15~mm icons with a droplet volume of 200~\textmu L, which provided a suitable compromise between load-carrying capability, resistance to tilting, and compatibility with direct manipulation. 

We note that present experiments employed a basic single-focus attractive actuation scheme without field optimization. Previous work has demonstrated that experimental hologram optimization can substantially enhance trajectory control with only a small number of iterations~\cite{Fushimi2024}. Such optimization is therefore expected to improve accuracy even in regimes that currently exhibit partial transport. Moreover, closed-loop feedback control based on real-time position sensing could be implemented to dynamically correct trajectory errors during operation. Incorporating the coupled icon–droplet geometry into hologram design using model-based approaches such as those reported by Inoue \textit{et al.}~\cite{Inoue2019} may further strengthen lateral force generation and stability. Nevertheless, the present results show that even this simple actuation scheme is sufficient to achieve stable transport of loaded icons across a broad parameter range, establishing the fundamental feasibility of liquid-mediated acoustic manipulation and providing a foundation for future system-level refinement.

\subsection{Dynamic Stability During Transport of Elevated Payloads}
\label{sec:dynamic_stability}

Building on the above findings, we next evaluated whether the proposed solid--fluid acoustic interface could support and transport externally mounted three-dimensional structures. We placed a 3D-printed tower on top of a Janus icon and translated the combined structure along a horizontal linear trajectory (Fig.~\ref{fig:moving_printed}). The tower had a height of 24~mm and a base diameter of 10~mm, and was fabricated using a Bambu Lab X1 Carbon printer using red PLA at a layer height of 0.2~mm. The total mass of the printed structure was 137~mg. The original STL model was obtained from Thingiverse\footnote{\url{https://www.thingiverse.com/thing:311002}} under a CC-BY-SA license.

During the experiment, the icon--object assembly was driven back and forth along a straight path following a sinusoidal trajectory with an oscillation frequency of 10~Hz, using the same mesh height and focal-point height as in the previous section. The structure remained stably supported and followed the prescribed motion as shown in Supplementary Video~1.

One point to note is the gradual spinning of the icon during translation, which can also be observed in experiments without an external structure mounted on top. This behavior may arise from asymmetries in the acoustic radiation force that locally deform the supporting droplet. When the icon is small and tilted, such rotational motion can cause externally mounted loads to slip or detach during transport. The effect may be further amplified by vertically extended structures, which increase sensitivity to small tilts by shifting the center of gravity. These observations indicate that active mitigation strategies, such as hologram optimization to rebalance the load, may be required in applications where load orientation and attachment must be maintained during transport.

\subsection{Tangible Interaction with Janus Icons}
\label{sec:tangible_interaction}
Tangible user interfaces, first articulated in the Tangible Bits framework by Hiroshi Ishii~\cite{Ishii1997}, established a research direction in human–computer interaction centered on embodied interaction and the coupling of digital information with physical form, a direction later deepened through the concept of Radical Atoms, which envisions dynamically reconfigurable physical representations of digital information~\cite{Ishii2012}. The Janus icon platform extends this lineage by introducing a liquid-mediated tangible substrate, in which a supporting water droplet provides compliant mechanical response without embedded actuators. In contrast to prior droplet-based systems such as electrowetting-on-dielectric or pyroelectric interfaces—where droplets are typically small, transient, and not designed for sustained human contact—the use of larger, mechanically stable droplets enables direct, repeatable touch interaction. Here, the liquid layer functions as a reconfigurable elastic medium that supports both manipulation and tactile input.

We demonstrate two representative interaction modalities. We evaluate button-like interaction by vertically compressing the icon, as shown in Fig.\ref{fig:interactiveicon}a. When pressed downward, the supporting droplet deforms and generates a restoring force from droplets, providing perceivable tactile pushback. The icon can be depressed until it contacts the underlying mesh. Upon release, the droplet reforms beneath the icon and restores it to its original position (Supplementary Video 2). This self-recovery behavior arises from asymmetric wettability: the hydrophobic mesh and top surface repel water and prevent submergence of the icon during compression, while the hydrophilic bottom surface promotes rewetting and reattachment. Even after substantial deformation, residual liquid on the underside facilitates droplet reformation, enabling repeated operation without mechanical springs or embedded actuators.

We further demonstrate dynamic vibration of the icon by exploiting droplet resonance. Natural oscillation modes of liquid droplets, described in classical capillary oscillation theory, provide a mechanism for generating periodic mechanical motion. By applying amplitude modulation to the phased-array output while maintaining a single acoustic focus beneath the icon, the droplet can be driven into oscillation (Supplementary Video~3), where the resulting vibration is sufficiently strong to be observed visually and to allow direct physical contact. Using the same mesh height and focal-point height as in previous experiments, we drove the droplet with amplitude-modulated acoustic excitation at 10~V and a water volume of 200~\textmu L, and swept the modulation frequency from 5 to 40~Hz in 1~Hz increments. The resulting vibration of the icon was measured using a laser Doppler vibrometer (Polytec PSV-500-3D-Xtra), with three repeated measurements per frequency to obtain the mean and standard deviation. As shown in Fig.~\ref{fig:interactiveicon}b, a clear resonance peak appears around 22~Hz, at which the oscillation amplitude becomes readily visible.

Comparison with the classical Rayleigh--Lamb prediction~\cite{Lamb1932} for a free, perfectly spherical droplet of equal volume (green solid line in Fig.~\ref{fig:interactiveicon}b) reveals substantial deviation from the measured peak. This discrepancy is expected, as the droplet in the present system is constrained by the mesh, coupled to the icon, and subject to asymmetric wetting conditions, all of which modify its equilibrium shape and restoring forces. Together, these results demonstrate that the liquid layer can function both as a compliant spring for quasi-static input and as a resonant mechanical element for dynamic actuation, enabling reconfigurable, passive, and electronics-free icon for tangible interaction.

\subsection{System-Level Performance, Limitations, and Opportunities for Improvement}
\label{sec:limitations_outlook}

The experimental results demonstrate that the proposed solid--fluid acoustic interface can support stable transport, dynamic manipulation, and tangible interaction across a wide range of operating conditions. At the same time, overall performance is governed by a highly coupled, nonlinear and multi-variable physical system, which makes precise numerical prediction and analytical optimization challenging. Transport and interaction behavior depend on interrelated factors including surface wettability, droplet volume, icon geometry, load distribution, acoustic field distribution, and environmental conditions. In practice, several of these parameters evolve over time, for example due to gradual changes in surface properties, water evaporation, or contamination, further complicating long-term stability.

In addition, system performance cannot be improved solely by increasing the driving voltage. Although higher acoustic amplitude can enhance radiation forces, excessive excitation may induce undesirable behaviors such as icon overturning. This imposes upper bounds on actuation strength and motivate the need for balanced field design rather than brute-force amplification.

Despite these challenges, the present results also suggest multiple directions for systematic improvement. First, alternative combinations of liquids and icon materials (such as silicone oil) could be explored to tailor surface tension, viscosity, and wetting contrast, potentially improving robustness and reducing sensitivity to environmental variations. Second, optimization of acoustic holograms and multi-focus field configurations may enable stronger lateral confinement and improved trajectory control while suppressing rotational drift and lateral deviation. Third, closed-loop control incorporating real-time sensing of icon position, orientation, and droplet state could substantially enhance stability and repeatability, particularly under dynamic loading and user interaction.

Further improvements may be achieved by integrating additional sensing modalities, such as optical or acoustic feedback, to monitor interface deformation and resonance states. Such sensing capabilities would enable adaptive modulation strategies that compensate for evaporation, surface aging, and payload variation. Finally, improved physical modeling of mesh-supported, asymmetrically wetted droplets will be essential for predictive design and scalable optimization.
\section{Conclusion}
This work presents PolygonWave, a solid--fluid acoustic interface based on asymmetrically wetted Janus icons that enables contactless transport, dynamic manipulation, and tangible interaction at the centimeter scale. By coupling programmable acoustic radiation forces with capillary-mediated support, the system allows computational intent to be expressed through stable, touchable artifacts without embedded actuators or direct mechanical coupling.

Through systematic characterization and interactive demonstrations, we showed that water droplets can function as mechanically active elements, providing compliance, self-recovery, and resonance-based feedback. These properties enable forms of tangible interaction that are difficult to achieve with existing approaches.

At the same time, performance emerges from coupled interactions among acoustic fields, wettability, geometry, and environmental conditions, highlighting both the potential and the challenge of solid--fluid acoustic actuation. Taken together, these results indicate that the present system represents an initial step toward a general-purpose solid--fluid acoustic tangible interface, with substantial opportunities for improving robustness, scalability, and autonomy through advances in materials, field synthesis, sensing, and control.
\section{Methods}
\threesubsection{Physical Properties and Preparation of Icons} Balsa wood was selected as the substrate owing to its extremely low density, intrinsic hydrophilicity, and ease of fabrication. Its naturally hydrophilic surface eliminates the need for chemical treatment on one side of the icon, allowing selective modification of only the opposing surface.\\
Commercially available 1~mm-thick balsa wood sheets were used. Each sheet was temporarily affixed to paper using tape such that only one surface was exposed during coating. The exposed surface was treated with a commercially available superhydrophobic coating (Ultra-Ever Dry, Ultratech International, Inc.). A bottom coat was first applied uniformly and allowed to dry fully according to the manufacturer’s specifications. A top coat was then applied and dried at ambient room temperature. The opposite surface was left untreated and retained its native hydrophilic character.\\
After surface treatment, the balsa wood sheets were cut and patterned using a laser cutter (xTool S1, 10~W laser diode module). Cutting was performed at 100\% laser output with a translation speed of 8~mm~s$^{-1}$. The superhydrophobic surface was oriented upward, and the laser was applied from the coated side through to the untreated side. This configuration enabled the engraving of concentric reference rings on the superhydrophobic surface while preserving the hydrophilic side. The engraved rings were used as alignment guides to position added masses as close as possible to the geometric center of each icon. Engraving was carried out at 24\% laser power with a head movement speed of 75~mm~s$^{-1}$. These parameters were empirically calibrated, as predefined material settings for 1~mm-thick balsa wood were not available in the xTool native material library. Laser cutting resulted in localized scorching along the sidewalls of the icons. 

After cutting, each icon was individually labeled and stored in plastic bags for storage to minimize contamination and moisture uptake. The mass of each icon was measured after packaging. The measured masses were 0.011~g, 0.021~g, and 0.030~g for icons with diameters of 12~mm, 15~mm, and 18~mm, respectively. All mass measurements in the manuscripts were performed using an analytical balance (Shimadzu TW423N) with a resolution of 0.001~g.

The untreated balsa wood surface exhibited strong hydrophilicity; upon placement of a sessile water droplet, the liquid fully spread over the surface, and a contact angle could not be measured. In contrast, the superhydrophobic coated surface exhibited a static water contact angle of 116$^\circ$.

\threesubsection{Physical Properties and Preparation of Mesh}
The mesh used in this study was a specially manufactured stainless-steel filter (SUS304H, hereafter referred to as the mesh). In our previous work, commercially available woven meshes were used~\cite{Koroyasu2023}; however, the woven structure occasionally led to nonuniform hydrophobic coating, particularly at the interlaced regions. To eliminate this source of variability, the mesh used in the present study was fabricated from a single stainless-steel sheet, thereby avoiding weaving-induced inhomogeneities. The mesh was manufactured by Ring Co., Ltd. (Yao, Osaka, Japan).

The mesh was fabricated from SUS304H stainless steel. The overall dimensions of the sheet were 200~mm $\times$ 180~mm with a thickness of 0.1~mm. Circular through-holes with diameters ranging from 0.39~mm to 0.51~mm were arranged in a staggered (triangular) pattern with a lattice angle of 60$^\circ$. The hole pattern was confined to a square region of size 140~mm $\times$ 140~mm located at the center of the sheet. The open-area ratio within the patterned region was 53\%, and the fabrication tolerance on the hole diameter was specified as $\pm$0.02~mm.

Both sides of the mesh were treated with the same commercially available superhydrophobic coating used for the icons (Ultra-Ever Dry, Ultratech International, Inc.). The bottom coat was first applied uniformly and allowed to dry fully according to the manufacturer’s instructions, followed by application of the top coat and drying at ambient room temperature. No additional surface treatments were applied. The resulting mesh was mechanically robust, planar, and free of woven junctions, providing a geometrically uniform substrate for subsequent experiments. Prior to use, the mesh was handled carefully to avoid mechanical deformation. After coating, the mesh exhibited superhydrophobic behavior, with a static water contact angle of 121$^\circ$.

\threesubsection{Phased Array Transducer}
The phased array transducer system used in this study was based on the OpenMPD design framework reported by Montano-Murillo \textit{et al.}~\cite{Montano-Murillo2023}. OpenMPD provides a low-level control architecture for multimodal particle-based displays and enables precise spatiotemporal modulation of acoustic fields.

The array consisted of 256 ultrasonic transducers (Tamura SA-40A1) operating at a nominal frequency of 40~kHz. The transducers were arranged in a $16 \times 16$ square lattice configuration. The transducer array was mounted in a custom holder fabricated using fused deposition modeling (FDM) 3D printing. A stainless-steel mesh was positioned parallel to the transducer surface at a distance of 126~mm from the array.

Communication with the transducer control board was established via a USB-C interface and operated using a computer running the Windows operating system. The system supported communication update rates of up to 10~kHz. Custom-written Python scripts were used to control the array and generate the desired acoustic driving signals. Unless otherwise stated, a supply voltage of 14.5~V was applied to the transducer array.

A single-focus acoustic hologram is obtained by assigning the phase delay
\begin{equation}
\phi_t = -\frac{2 \pi f_0}{c_0} \left[ d(\vb*{x_f}, \vb*{x_t}) - d(0, \vb*{x_f}) \right],
\end{equation}
where $f_0 = 40$ kHz and $c_0 = 341~\mbox{ms}^{-1}$ denote the operating frequency and the speed of sound in air, respectively. The function $d(\cdot,\cdot)$ represents the Euclidean distance between two points. Here, $\vb*{x_f}$ is the focal position and $\vb*{x_t}$ denotes the transducer locations. 

\threesubsection{Use of AI assisted services}
We used ChatGPT 5.2 for image generation to stylize the images in Figure 1. The generated images were subsequently vectorized using the Image Trace function in Autodesk Illustrator. Code editing was performed using Windsurf, and language editing was assisted by ChatGPT 5.2. All outputs were manually reviewed by the authors to ensure accuracy and legitimacy.

\medskip

\textbf{Supporting Information} \par 
Supporting Information is available from the Wiley Online Library or from the author.

\textbf{Data Availability Statement} \par 
The data that supports the findings of this study are available in the supplementary material of this article.

\textbf{Conflict of Interest} \par 
The authors have no competing interests to disclose.

\medskip
\textbf{Acknowledgements} \par 
This work was supported by JSPS KAKENHI Grant Number JP23K16916.

\medskip

%

\bibliographystyle{MSP}
\bibliography{output}

\begin{figure}
\centering
  \includegraphics[width=\linewidth]{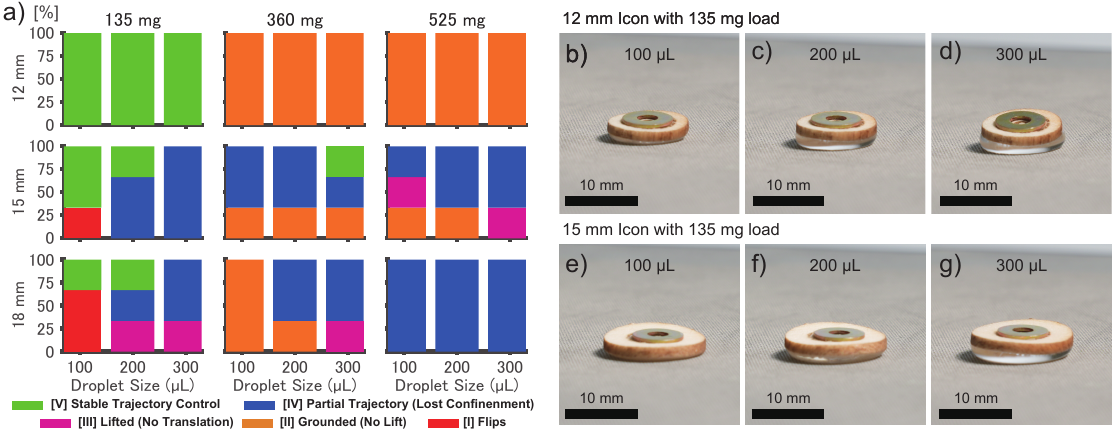}
  \caption{Transport performance across icon diameter, droplet volume, and applied load. 
(a) Summary of transport outcomes for combinations of icon size, droplet volume, and load. Each bar indicates the proportion of trials classified as stable trajectory control [V], partial trajectory [IV], lifted without translation [III], grounded without lift [II], or flipped [I]. 
(b--d) Side-view images of 12~mm icons with increasing droplet volumes (100, 200, and 300~\textmu L), showing progressive vertical elongation and increased tilt. 
(e--g) Corresponding images for 15~mm icons, demonstrating more uniform droplet distribution and improved mechanical stability over the same volume range.}
\label{fig:boat1}
  \label{fig:boat1}
\end{figure}

\begin{figure}
\centering
  \includegraphics[width=0.5\linewidth]{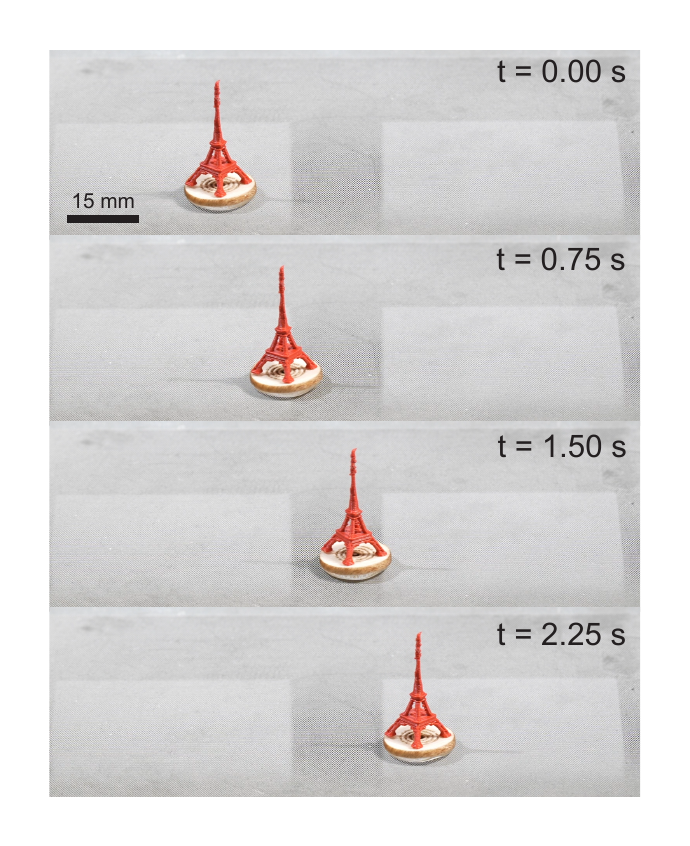}
  \caption{Dynamic stability during transport of an elevated payload. 
A 24~mm-tall, 10~mm-base 3D-printed tower (137~mg, red PLA) mounted on a Janus icon is translated along a horizontal linear trajectory. The assembly follows a sinusoidal back-and-forth motion at 10~Hz while maintaining capillary support and translational stability. Despite its elevated center of mass, the structure remains stably coupled to the supporting droplet throughout motion (see Supplementary Video~1).}
\label{fig:moving_printed}
\end{figure}

\begin{figure}
\centering
  \includegraphics[width=0.5\linewidth]{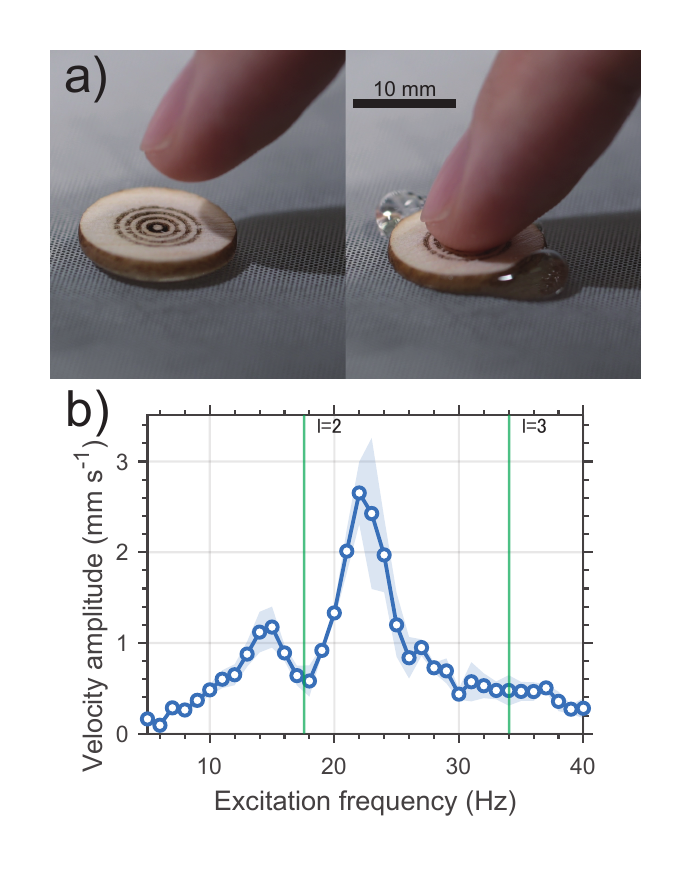}
\caption{Tangible interaction enabled by liquid-supported Janus icons. 
(a) Button-like input through vertical compression of the icon, in which the supporting droplet provides compliant tactile feedback and self-recovery after release (see supplementary video 2).
(b) Vibration generated by droplet resonance (see supplementary video 3). Measured vibration amplitude of the icon (15 mm) as a function of modulation frequency for a 200~\textmu L droplet at 10~V, showing a clear resonance peak around 22~Hz. The solid green lines indicate the classical Rayleigh--Lamb predictions for the $l=2$ and $l=3$ capillary modes of a free spherical droplet of equal volume.}
\label{fig:interactiveicon}
\end{figure}

\end{document}